\documentclass[reprint, amsmath, amssymb, aps, superscriptaddress]{revtex4-1}
\usepackage{format}
\allowdisplaybreaks

\begin{document}

\preprint{APS/123-QED}

\title{ Thermal Conductivity of an Ultracold Paramagnetic Bose Gas  }

\author{Reuben R. W. Wang}
\author{John L. Bohn}
\affiliation{JILA, University of Colorado, Boulder, Colorado 80309, USA}

\date{\today} 

\begin{abstract}

We analytically derive the transport tensor of thermal conductivity in an ultracold, but not yet quantum degenerate, gas of Bosonic lanthanide atoms using the Chapman-Enskog procedure. The tensor coefficients inherit an anisotropy from the anisotropic collision cross section for these dipolar species, manifest in their dependence on the dipole moment, dipole orientation, and $s$-wave scattering length. These functional dependencies open up a pathway for control of macroscopic gas phenomena via tuning of the microscopic atomic interactions. As an illustrative example, we analyze the time evolution of a temperature hot-spot which shows preferential heat diffusion orthogonal to the dipole orientation, a direct consequence of anisotropic thermal conduction.

\end{abstract}

\maketitle

\section{\label{sec:introduction} Introduction}

Ultracold gases of spin-polarized magnetic atoms, such as dysprosium or erbium, have led to a wealth of novel phenomena in the quantum degenerate regime, as reviewed recently in Ref.~\cite{Chomaz22_arxiv}. Far less studied is the regime of such gases just above the temperature of quantum degeneracy. In this regime, a small magnetic field can assure that the atoms remain polarized, whereby the classical fluid equations of motion inherit anisotropy due to this polarization.  In particular, the transport coefficients -- the thermal conductivity and the viscosity -- inherit an anisotropy from the microscopic collision dynamics of the scattering dipoles.

In certain cases, the results of this collisional anisotropy are well known.  They have already been shown to result in anisotropic thermalization in normal phase  ultracold gases, and can be used as a tool for measuring scattering lengths \cite{Sykes15_PRA, Tang16_PRL, Wang20_PRA, Wang21_PRA,Tang15_PRA, Maier15_PRA, Lucioni18_PRA, Durastante20_PRA, Patscheider21_arxiv}. These experiments have been modeled using perturbation theory around the equilibrium Boltzmann distribution of a gas, an analysis that has proven highly successful.
Following on such success, it seems worthwhile to present the systematic derivation of the continuum fluid equations of motion for the ultracold paramagnetic gas.  The present paper takes the first step in this program, by deriving the anisotropic thermal conductivity tensor from the differential cross section in dipolar lanthanide gases \cite{Bohn14_PRA}. 
This is done by means of the Chapman-Enskog formalism \cite{Chapman90_CUP}, leading to density independent coefficients valid in the dilute regime \cite{Hanley72_Phys}. 

We focus here on Bosonic samples, which also offer a quantum mechanical $s$-wave scattering length $a_s$ \cite{Mott49_OUP}, tunable via a multitude of Fano-Feshbach resonances.  Thus the anisotropy of the heat conduction tensor is under direct experimental control.  We note that our results here are unlike studies where anisotropic transport tensors arise due to internal degrees of freedom or long-ranged interactions \cite{Dehkordi12_JPCS}, such as in systems of dilute plasmas \cite{Braginskii63_RPP, Daybelge70_JAP, Bruno06_POP, deGroot13_DP} and ferrofluids \cite{Saluena92_JCP, Suh15_MT}.

The remainder of this manuscript is organized as follows: 
In Sec.~\ref{sec:ChapmanEnskog} and \ref{sec:thermal_conductivity}, we analytically derive the anisotropic transport tensor of thermal conductivity emergent from dipolar collisions.  
The continuum conservation equations are introduced in Sec.~\ref{sec:fluid_dynamics}, permitting a model for fluid dynamic studies in ultracold gases. This model is used to study a simple experimental scenario of thermal diffusion of a temperature hot-spot in Sec.~\ref{sec:hotspot_dispersion}. Finally, discussions and concluding remarks are drawn in sec.~\ref{sec:conclusions}.

\section{ \label{sec:ChapmanEnskog} The Chapman-Enskog Procedure }

The study of transport phenomena is mature and extensive, having applications to all fields of science and engineering \cite{Bird06_Wiley, Bottin06_PAS, Plawsky09_CRC, Truskey10_P}. Central to the analysis of transport are the equations of conservation and constitution, which describe the dynamics of state variables (e.g. mass, flow-velocity and energy) and their response to external stimuli. If only weakly perturbed, the response of a system is completely described by linear constitutive relations and the associated, medium-specific, transport coefficients.  

In the present context, we consider an ultracold, dilute gas of Bosonic lanthanide atoms, in their spin-stretched ground state and in a sufficient magnetic field that they remain in this ground state in spite of collisions. The gas is then paramagnetic, with a preferred spatial axis  determined by the field direction.  Moreover, we explicitly consider only temperatures above the critical temperature of Bose-Einstein condensation, whereby the thermodynamics of the gas is governed by Maxwell-Boltzmann statistics. While we focus on magnetic atoms here, the results should of course be applicable to ultracold gases of polar molecules. 

\begin{figure}[ht]
    \centering
    \includegraphics[width=0.55\columnwidth]{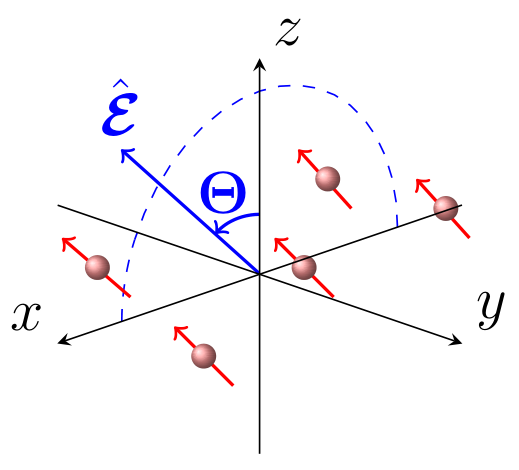}
    \caption{ A visualization of dipoles (red) aligned with an external field along the dipole-alignment axis, $\hat{\boldsymbol{{\cal E}}}$ (blue), in the laboratory coordinate frame. }
    \label{fig:lab_frame_axes}
\end{figure}

In such a gas, local equilibrium occurs by means of dipolar collisions parameterized by the scattering length $a$, and magnetic dipole length $a_d = C_\mathrm{dd} m / ( 8 \pi \hbar^2 )$, where $C_{\text{dd}} = \mu_0\mu^2$ ($\mu_0$ is the vacuum permeability). We take that all the dipoles are aligned along a dipole-alignment axis $\hat{\boldsymbol{{\cal E}}}$, by means of a large external field taken to lie in the $x,z$-plane (illustrated in Fig.~\ref{fig:lab_frame_axes}). We thus envision experiments conducted in a fixed frame of reference, with the polarization orientation free to be tuned relative to this axis.

Close to local thermal equilibrium, re-equilibration processes are encapsulated by transport coefficients (e.g. viscosity, thermal conductivity, etc) derivable from a microscopic picture by methods established by Chapman and Enskog \cite{Chapman90_CUP}. The development we present here close follows that of \cite{Bond65_AW}.

Within length scales on the order of the atomic mean-free path, atomic interactions are dominated by collisional processes. The local distribution of atoms in flow thus has dynamics well described by the Boltzmann transport equation \cite{Huang63_NY, Reif09_Waveland}
\begin{subequations}
\begin{align} \label{eq:Boltzmann_equation}
    & \left( \frac{ \partial }{ \partial t } + v_i \partial_i \right) f(\boldsymbol{r}, \boldsymbol{v}) = {\cal C}[ f(\boldsymbol{r}, \boldsymbol{v}) ], \\
    & {\cal C}[ f ] = \int d\Omega' \frac{d\sigma}{d\Omega'} \int d^3 v_1 \abs{\boldsymbol{v} - \boldsymbol{v}_1} \left( f' f_1' - f f_1 \right),
\end{align}
\end{subequations}
where $f(\boldsymbol{r}, \boldsymbol{v})$ is the phase space distribution function and ${\cal C}[f]$ is the two-body collision integral. We adopt the convention that all repeated indices are summed over unless otherwise specified, and primes denote post-collision velocities for pairs of atoms colliding with incoming velocities $\boldsymbol{v}$ and $\boldsymbol{v}_1$. We also adopt the compact notation $f_1 = f(\boldsymbol{r}, \boldsymbol{v}_1)$ and $f' = f(\boldsymbol{r}, \boldsymbol{v}')$. At thermal equilibrium (denoted by subscript $0$), the gas has number density $n_0 = \rho_0 / m$, only dependent on temperature $n_0 = n_0(\beta)$, and velocities that are Boltzmann distributed 
\begin{align} \label{eq:equilibrium_ansatz}
    f_0(\boldsymbol{u}, \beta) &= n_0(\beta) c_0(\boldsymbol{u}, \beta) \nonumber\\
    &= n_0(\beta) \left( \frac{ m \beta }{ 2 \pi } \right)^{3/2} \exp\left( - \frac{ m \beta }{ 2 } \boldsymbol{u}^2 \right),
\end{align}
where $\beta = (k_B T)^{-1}$, $\boldsymbol{u}^2 = u_k u_k$, and $\boldsymbol{u}(\boldsymbol{r}) = \boldsymbol{v} - \boldsymbol{U}(\boldsymbol{r})$ is the molecular velocity relative to the flow velocity, also called peculiar velocity. In close-to-equilibrium scenarios, we can consider the out-of-equilibrium atomic distribution to take the form
\begin{align} \label{eq:close2equilibrium_ansatz}
    & f(\boldsymbol{r}, \boldsymbol{u}, \beta) \approx f_0(\boldsymbol{u}, \beta) [ 1 + \Phi(\boldsymbol{r}, \boldsymbol{u}, \beta) ], 
\end{align}
with a perturbation function $\Phi$, that must satisfy 
\begin{subequations} \label{eq:ansatz_conservation_laws}
\begin{align}
    & \int d^3 u f_0(\boldsymbol{u}) \Phi(\boldsymbol{r}, \boldsymbol{u}, \beta) m = 0, \\
    & \int d^3 u f_0(\boldsymbol{u}) \Phi(\boldsymbol{r}, \boldsymbol{u}, \beta) m \boldsymbol{u} = 0, \\
    & \int d^3 u f_0(\boldsymbol{u}) \Phi(\boldsymbol{r}, \boldsymbol{u}, \beta) \frac{1}{2} m \boldsymbol{u}^2 = 0,
\end{align}
\end{subequations}
as a result of mass, momentum and energy conservation respectively. Enskog's prescription of successive approximations then renders the Boltzmann equation, to leading non-trivial order, as
\begin{align} \label{eq:1stOrder_BoltzmannEquation}
    \left( \frac{\partial }{ \partial t } + v_i \partial_i \right) f_0 \approx C[ f_0 \Phi ].
\end{align}
Physically, this approximation is motivated by establishing a separation of scales between phenomena of interest. We are concerned with the regime in which macroscopic fluid dynamics is governed by length scales $\lambda$ (e.g. wavelengths) much larger than the mean-free path $L$, of its constituent atoms (i.e. the regime of small Knudsen number Kn $ = { L \lambda^{-1} } \ll 1$). Furthermore, the period over which such dynamics occurs is much longer than the timescales associated to collisions. Therefore, Eq.~(\ref{eq:1stOrder_BoltzmannEquation}) effectively makes an adiabatic approximation that separates the macro and micro-scale phenomena.  We refer to the fluid dynamics as ocurring on ``macro-scales", whereas collisional interactions are said to occur on ``micro-scales".

Under the approximation described above, the left-hand side of Eq.~(\ref{eq:1stOrder_BoltzmannEquation}) evaluates to
\begin{align} \label{eq:Boltzmann_LHS}
    & \left( \frac{\partial }{ \partial t } + v_k \partial_k \right) f_0  
    = f_0 \big[ V_{k} \partial_k ( \ln T ) + m \beta W_{k\ell} D_{k\ell} \big]. 
\end{align}
where 
\begin{subequations}
\begin{align}
    & V_i(\boldsymbol{u}) \equiv \left( \frac{ m \beta \boldsymbol{u}^2 }{ 2 } - \frac{ 5 }{ 2 } \right) u_i, \\
    & W_{ij}(\boldsymbol{u}) \equiv u_i u_j - \frac{ 1 }{ 3 } \delta_{ij} \boldsymbol{u}^2, \\
    & D_{ij}(\boldsymbol{U}) \equiv \frac{ 1 }{ 2 } \left( \partial_{j} U_i + \partial_i U_j \right) - \frac{ 1 }{ 3 } \delta_{ij} \partial_k U_k.
\end{align}
\end{subequations}
The derivation of this result is detailed in App.~\ref{app:1st_order_ChapmanEnskog}. The collision integral on the right-hand side of Eq.~(\ref{eq:1stOrder_BoltzmannEquation}) is then
\begin{align}
    C[ f ] \approx \int d^3 u_1 & \abs{ \boldsymbol{u} - \boldsymbol{u}_1 } f_0(\boldsymbol{u}) f_0(\boldsymbol{u}_1) \int d\Omega' \frac{ d \sigma }{ d \Omega' } \Delta \Phi, \label{eq:collision_integral}
\end{align}
where $\Delta \Phi = \Phi' + \Phi'_1 - \Phi - \Phi_1$. Since Eq.~(\ref{eq:collision_integral}) is linear in $\Phi$, and Eq.~(\ref{eq:Boltzmann_LHS}) is linear in the quantities $\partial_i \ln T$ and $\partial_j U_i$, one can infer an {\it ansatz} for the scalar function $\Phi$, of the form 
\begin{align} \label{eq:out_of_equilibrium_ansatz} 
    \Phi(\boldsymbol{u}, \beta) &= {\cal B}_k \partial_k ( \ln T ) + m \beta {\cal A}_{k\ell} D_{k\ell},
\end{align}
where $\boldsymbol{{\cal B}}$ (vector) and $\boldsymbol{{\cal A}}$ (2-rank tensor) are functions of $\boldsymbol{u}$ and $\beta$. The {\it ansatz} above allows a separation of Eq.~(\ref{eq:1stOrder_BoltzmannEquation}) into an equation in velocity  gradients, and those in temperature gradients:
\begin{subequations}
\begin{align}
    & f_0 \: W_{k\ell} D_{k\ell} \approx C[ f_0 {\cal A}_{k\ell} ] D_{k\ell}, \label{eq:viscosity_BE} \\
    & f_0 \: V_k \partial_k ( \ln T ) \approx C[ f_0 {\cal B}_k ] \partial_k ( \ln T ), \label{eq:conductivity_BE}
\end{align}
\end{subequations}
which upon comparing terms, further motivate $\boldsymbol{{\cal B}}$ and $\boldsymbol{{\cal A}}$ to be written as
\begin{subequations}
\begin{align}
    & {\cal A}_{ij}(\boldsymbol{u}, n_0, \beta) = W_{k\ell}(\boldsymbol{u}) a_{k \ell i j}(u, n_0, \beta), \\
    & {\cal B}_{i}(\boldsymbol{u}, n_0, \beta) = V_j(\boldsymbol{u}) b_{j i}(u, n_0, \beta),
\end{align}
\end{subequations}
where $u = \abs{\boldsymbol{u}}$, and the coefficients $a_{k \ell m n}(u, n_0, \beta)$ and $b_{k \ell}(u, n_0, \beta)$ are introduced as variational {\it ansatz}. These variational coefficients can, in general, be expressed as an infinite linear combination of Sonine polynomials (a.k.a. associated Laguerre polynomials). The assumption of a low temperature gas however, allows us to approximate $\boldsymbol{a}$ and $\boldsymbol{b}$ with only the first term in the summation series, which is $u$-independent. Such an approximation has been shown to give good accuracy (relative errors of $\sim 1 \%$) in computing transport coefficients for gases of isotropic scatterers \cite{Pekeris57_PNASUSA, Loyalka07_PA, Reif09_Waveland}. We are thus left with  
\begin{align} 
    \Phi(\boldsymbol{u}, \beta) &= V_{\ell}(\boldsymbol{u}) b_{\ell k}(n_0, \beta) \partial_k ( \ln T ) \nonumber\\
    &\quad\quad + m \beta W_{i j}(\boldsymbol{u}) a_{i j k \ell}(n_0, \beta) D_{k\ell}.
\end{align}
The coefficient $a$ and $b$ are determined for a particular gas by the microscopic scattering theory of the constituents, a task to which we now turn.

\section{ \label{sec:thermal_conductivity} Thermal Conductivity in Dipolar Gases }

Thermal conduction in a dilute gas arises through a transfer of kinetic energy by kinetic transport of the gaseous atoms, out of a region of fluid, resulting in a heat flux \cite{deGroot13_DP}
\begin{align} \label{eq:heat_flux}
    J_i(\boldsymbol{r}, t) = \int d^3 u  f(\boldsymbol{r}, \boldsymbol{u}, t) \frac{1}{2} m \boldsymbol{u}^2 u_i. 
\end{align}
For a first-order approximation, we adopt the {\it ansatz} of Eq.~(\ref{eq:out_of_equilibrium_ansatz}) to compute the integral above. The $\boldsymbol{{\cal A}}$ associated term does not contribute to the heat flux integral, leaving us with
\begin{align} \label{eq:heat_flux_transport}
    J_i &= \frac{m}{2} \int d^3 u  f_0(\boldsymbol{u}) [ 1 + \Phi(\boldsymbol{u}) ] \boldsymbol{u}^2 u_i \nonumber\\ 
    &= \left( \frac{ k_B m \beta }{ 2 } \int d^3 u \: f_0(\boldsymbol{u}) \boldsymbol{u}^2 u_i V_k b_{k j} \right) { \partial_j T }, 
\end{align}
where the local temperature $T(\boldsymbol{r},t)$ is written in terms of its kinetic definition,
\begin{align} \label{eq;kinetic_temperature_definition}
    \frac{3}{2} k_B T = \frac{ 1 }{ n(\boldsymbol{r}, t) } \int d^3 u \: f(\boldsymbol{r}, \boldsymbol{u}, t) \frac{1}{2} m \boldsymbol{u}^2.
\end{align}

Additionally, we say that this flow of kinetic energy occurs across a temperature gradient via Fourier's law of heat conduction 
\begin{align} \label{eq:phenom_heat_flux}
    J_i = -\kappa_{ij} \partial_j T,
\end{align} \vspace{5pt}
where $\boldsymbol{\kappa}$ is the thermal conductivity, a 2-rank tensor. A comparison of Eq.~(\ref{eq:heat_flux_transport}) and Eq.~(\ref{eq:phenom_heat_flux}), then tells us that the thermal conductivity is found via the integral
\begin{align}
    \kappa_{ij} &= - \left( \frac{ k_B m \beta }{ 2 } \int d^3 u \: f_0(\boldsymbol{u}) \boldsymbol{u}^2 u_i V_k \right) b_{k j} \nonumber\\ 
    &= -\frac{ 5 n_0 k_B }{ 2 m \beta } b_{i j}, 
\end{align}
assuming knowledge of the coefficients $b_{kj}$. 

The transport of kinetic energy across a temperature gradient is brought about by the flow of atoms mediated by collisions, allowing use of the Boltzmann equation to derive $\boldsymbol{b}(u)$, with the first-order Chapman-Enskog expansion.  Referring back to Eq.~(\ref{eq:conductivity_BE}), one finds that it is formally mathematically inconsistent but holds in an average sense over the atomic distribution by multiplying Eq.~(\ref{eq:conductivity_BE}) by $V_i(\boldsymbol{u}) d^3 u$, and integrating. This gives 
\begin{align}
    & \left( \int d^3 u \: f_0( \boldsymbol{u} ) V_i(\boldsymbol{u}) V_j(\boldsymbol{u}) \right) \partial_j ( \ln T ) \nonumber\\
    &\quad\quad\quad \approx \left( \int d^3 u \: V_i(\boldsymbol{u}) C[ f_0 V_k ] \right) b_{k j} \partial_j ( \ln T ),
\end{align} 
whereby the coefficients of $\partial_j ( \ln T )$ satisfy the relation
\begin{subequations} \label{eq:collision_integral_Nmatrix}
\begin{align}
    & N_{i k} b_{k j} = \delta_{ij}, \\
    \text{where}\quad & N_{i k} \equiv \frac{ 2 m \beta }{ 5 n_0 } \int d^3 u \: V_i C[ f_0 V_k ].
\end{align}
\end{subequations}
Finally, we evaluate these integrals incorporating the differential cross section for dipoles in the App.~\ref{app:collision_integral_T}, following the successful method developed in \cite{Wang20_PRA}. We cast the result in terms of the dimensionless functions
\begin{align}
    {\cal N}_{ij} = 
    \frac{ 1 }{ a^2 n_0 } \sqrt{ \frac{ m \beta }{ \pi } } N_{ij},
\end{align}
with
\begin{widetext}
\begin{subequations} \label{eq:calNij_terms}
\begin{align} 
    & {\cal N}_{11} = -\frac{ 256 }{ 15 } + \frac{ 256 a_d (3 \cos (2 \Theta ) - 1) }{ 225 a } - \frac{512 a_d^2 (3 \cos (2 \Theta ) + 13)}{ 4725 a^2 },
    \\
    & {\cal N}_{13} = -\frac{ 256 a_d \sin (2 \Theta ) }{ 75 a } + \frac{ 512 a_d^2 \sin (2 \Theta ) }{ 1575 a^2 },
    \\
    & {\cal N}_{22} = -\frac{ 256 }{ 15 } + \frac{ 512 a_d }{ 225 a } - \frac{ 8192 a_d^2 }{ 4725 a^2 },
\end{align}
\end{subequations}
\end{widetext}
with the additional relations
\begin{subequations}
\begin{align}
    & {\cal N}_{33}(\Theta) = {\cal N}_{11}(\Theta - \pi/2), \\
    & {\cal N}_{12}(\Theta) = {\cal N}_{23}(\Theta) = 0, \\
    & {\cal N}_{ij}(\Theta) = {\cal N}_{ji}(\Theta).
\end{align}
\end{subequations}
It then follows that the thermal conductivity tensor is given as
\begin{align} \label{eq:unitfree_coefficients}
    \boldsymbol{\kappa}(\Theta) 
    &= -\frac{ 5 k_B }{ 2 a^2 \sqrt{ \pi m \beta } } \boldsymbol{ {\cal N} }^{-1}(\Theta) \\
    &= -\frac{ 5 k_B }{ 2 a^2 \sqrt{ \pi m \beta } } \begin{pmatrix}
        \frac{ {\cal N}_{33} }{ {\cal N}_{11} {\cal N}_{33} - {\cal N}_{13}^2 } & 0 & \frac{ {\cal N}_{13} }{ {\cal N}_{13}^2 - {\cal N}_{11} {\cal N}_{33} } \\
        0 & \frac{ 1 }{ {\cal N}_{22} } & 0 \\
        \frac{ {\cal N}_{13} }{ {\cal N}_{13}^2 - {\cal N}_{11} {\cal N}_{33} } & 0 & \frac{ {\cal N}_{11} }{ {\cal N}_{11} {\cal N}_{33} - {\cal N}_{13}^2 } \\
    \end{pmatrix}. \nonumber
\end{align}
The structure of the tensor above along with Eq.~(\ref{eq:phenom_heat_flux}), imply that a temperature gradient along $x$ could result in a thermal flux along $z$, and vice versa. In the event that the dipoles are aligned along ${\hat z}$, that is $\Theta=0$, the Cartesian axes are the principal axes of $\boldsymbol{\kappa}$.  This situation leaves us with only two unique, nontrivial thermal conductivities $\kappa_{xx} = \kappa_{yy} \neq \kappa_{zz}$. 

\begin{figure}[ht]
    \centering
    \includegraphics[width=\columnwidth]{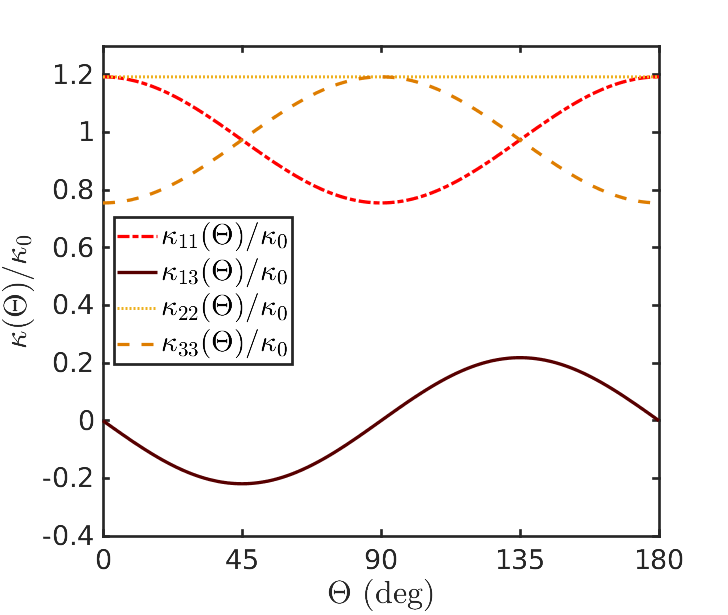}
    \caption{ The unit-free thermal conductivity tensor elements $\boldsymbol{\kappa} / \kappa_0$, as a function of the dipole-alignment angle $\Theta$, as defined in Eq.~(\ref{eq:unitfree_coefficients}) for native $^{164}$Dy ($a = 92$ a$_0$). The tensor elements $\kappa_{11} / \kappa_0$ (solid, dark red), $\kappa_{13} / \kappa_0$ (dot-dashed, red) and $\kappa_{22} / \kappa_0$ (dashed, orange) display a sinusoidal $\Theta$ dependence, whereas $\kappa_{33} / \kappa_0$ (dotted, yellow) is $\Theta$-independent due to the coordinate frame definition. The parameters considered here are for $^{164}$Dy with $a_d/a \approx 199/92$, taken from Ref.~\cite{Tang15_PRA}. 
    }
    \label{fig:kappas_vs_Theta}
\end{figure}

We plot in Fig.~\ref{fig:kappas_vs_Theta}, the coefficients of Eq.~(\ref{eq:unitfree_coefficients}) with values normalized by the isotropic coefficient $\boldsymbol{\kappa} / \kappa_0$ 
\footnote{ The coefficient $\kappa_0$ is exactly the result of Chapman and Enskog \cite{Chapman90_CUP}, but modified to include the quantum mechanical scattering length instead of a classical hardsphere radius. }, where
\begin{align} 
    \kappa_0 = \frac{ 75 k_B }{ 256 r_\mathrm{eff}^2 \sqrt{\pi m \beta} },
\end{align}
where $r_\mathrm{eff}^2 = 2 a^2 + 8 a_d^2 / 45$ is an effective isotropic radius obtained from an angular average of the dipolar differential cross section. The coefficients are plotted with the scattering and dipole lengths of native $^{164}$Dy ($a = 92 a_0$ and $a_d = 199 a_0$, where $a_0$ is the Bohr radius) \cite{Tang15_PRA}, which showcases the functional dependence on the angle $\Theta$ between the polarization and the laboratory $z$-axis. 

\section{ \label{sec:fluid_dynamics} Equations of Motion }

Having derived the transport tensor of thermal conductivity, macroscopic gas dynamics can now be studied under the lens of a continuum fluid formulation. The dynamics of fluids is characterized by spatial and temporal variations of macroscopic quantities such as the fluid mass density $\rho$, flow-velocity $\boldsymbol{U}$, and temperature $T$. These are related to the microscopic phase space distribution according to  
\begin{subequations}
\begin{align}
    & \rho(\boldsymbol{r}, t) = \int d^3 v f(\boldsymbol{r}, \boldsymbol{v}, t) m, \\
    & \boldsymbol{U}(\boldsymbol{r}, t) = \frac{1}{n(\boldsymbol{r}, \boldsymbol{v}, t)} \int d^3 v f(\boldsymbol{r}, \boldsymbol{v}, t) \boldsymbol{v}, \\
    & T(\boldsymbol{r}, t) = \frac{ 2 }{ 3 n(\boldsymbol{r}, \boldsymbol{v}, t) k_B } \int d^3 v f(\boldsymbol{r}, \boldsymbol{v}, t) \frac{1}{2} m \boldsymbol{v}^2. 
\end{align}
\end{subequations}
The associated hydrodynamic phenomena are well modeled, even in ultracold systems \cite{Nikuni98_JLTP}, by the continuum conservation equations \cite{deGroot13_DP} 
\begin{subequations} \label{eq:continuum_conservation_laws}
\begin{align}
    & \frac{ \partial \rho }{ \partial t } + \partial_j \left( { \rho U_j } \right) = 0, \\
    & \frac{\partial}{\partial t} \left( \rho U_i \right) + \partial_j \left( \rho U_j U_i \right) = \partial_j \sigma_{ij}, \\ 
    & \frac{\partial}{\partial t} (\rho T) + \partial_j \left( \rho T U_j \right) = \frac{ 2 m }{ 3 k_B } \left( \sigma_{ij} \partial_j U_i - \partial_j J_j \right), \label{eq:energy_conservation} 
\end{align}
\end{subequations}
where $\partial_i$ denotes a derivative with respect to coordinate ${r}_i$ ($i = 1,2,3$), and $m$ is the atomic mass. These equations are, in order, referred to as the continuity, Navier-Stokes and energy balance equations. As we have seen in the previous section, atom-atom collisions in the gas result in thermal transport and viscous effects, included into Eqs.~(\ref{eq:continuum_conservation_laws}) via the heat flux vector $J_j$, and pressure tensor \cite{Jog15_CUP}
\begin{subequations}
\begin{align}
    & \sigma_{ij} = -P \delta_{ij} + \tau_{ij}, \label{eq:stress_tensor_definition} \\
    & \tau_{i j} = \mu_{i j k \ell} \partial_{\ell} U_k,
\end{align}
\end{subequations}
where $P$ the thermodynamic pressure, $\tau_{ij} = \mu_{i j k \ell}$ the viscous stress tensor and $\mu_{i j k \ell}$ is the viscosity tensor. For the time being, we focus on the influence of thermal conductivity by assuming that all second derivatives of the flow-velocity are small, effectively rendering the viscous stress terms negligible (i.e. $\tau_{ij} \approx 0$). Consideration of the anisotropic viscosity is left to future work.

\section{ \label{sec:hotspot_dispersion} Diffusion of a Hot-Spot }

\begin{figure}[ht]
    \centering
    \includegraphics[width=\columnwidth]{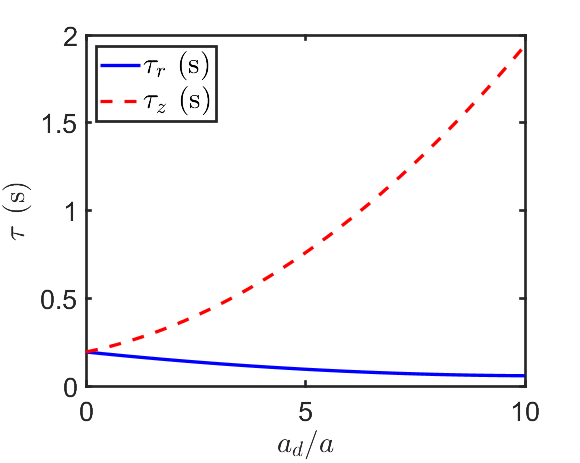}
    \caption{ Thermal relaxation time scales $\tau_r$ and $\tau_z$, vs the reduced dipole length $a_d / a$. The axial time scale $\tau_z$ is seen to be drastically larger than the radial time scale $\tau_r$ for large values of the reduced dipole length. }
    \label{fig:dispersion_timescale}
\end{figure}

As an example of anisotropy due to the thermal conductivity tensor, we consider a simple uniform gas experiment where a localized temperature hot-spot is induced, for example by heating the gas locally with a focused laser, then allowed to diffuse. For simplicity, we assume that the temperature field is excited perturbatively so that the temperature dynamics is described by its deviation from the uniform background temperature $T_0$, $T(\boldsymbol{r},t) = T_0\left[ 1 + \epsilon(\boldsymbol{r},t) \right]$. This permits a linearization of Eq.~(\ref{eq:energy_conservation}) to first-order in $\epsilon$, which gives
\begin{align}
    \frac{ \partial \epsilon }{ \partial t } \approx - \frac{ 2 }{ 3 } \partial_j U_j + \frac{ 2 }{ 3 n_0 k_B } \kappa_{ij} \partial_i \partial_j \epsilon.
\end{align}
At the onset of the hot-spot, the flow velocity ${\bf U}$ is taken as negligible, thus rendering the heat equation as 
\begin{align} \label{eq:diffusion_equation}
    \frac{\partial \epsilon}{\partial t} = {\cal D}_{ij} \partial_i \partial_j \epsilon,
\end{align}
in terms of a thermal diffusivity tensor
\begin{align}
    {\cal D}_{ij} \equiv \frac{2 }{3 n_0 k_B}\kappa_{ij}. 
\end{align}
We model the initial hot-spot as described by a Gaussian of width $\sigma$,
\begin{align}
    \epsilon(\boldsymbol{r}, t=0) = \epsilon_0 e^{ -\frac{ r^2 }{ 2 \sigma^2 } }.
\end{align}
Utilizing a Fourier expansion, one obtains the time-dependent solution to Eq.~(\ref{eq:diffusion_equation})
\begin{align}
    \epsilon(\boldsymbol{r}, t) &= \epsilon_0 \int \frac{ \sigma^3 d^3 K }{ (2 \pi)^{3/2} } e^{ -\frac{1}{2} K^2 \sigma^2 } e^{ -( \boldsymbol{K}^T \boldsymbol{{\cal D}} \boldsymbol{K} ) t } e^{ i \boldsymbol{K} \cdot \boldsymbol{r} }.  
\end{align}
The integral above can be evaluated analytically to give
\begin{subequations} \label{eq:hotspot_solution}
\begin{align}
    & \epsilon(\boldsymbol{r}, t) = \frac{ \epsilon_0 \sigma^3 }{ \sqrt{ 8 \det\left( \boldsymbol{M} \right) } } \exp\left( -\frac{ \boldsymbol{r}^T \boldsymbol{M}^{-1} \boldsymbol{r} }{ 4 } \right), \\
    & \boldsymbol{M} \equiv \frac{1}{2} \sigma^2 \mathbb{I} + \boldsymbol{{\cal D}} t, 
\end{align}
\end{subequations}
where $\mathbb{I}$ is the identity matrix. The solution above is further simplified if we assume that the dipoles define the $z$-axis, which is done here without loss of generality. The diffusion tensor is now diagonal with only 2 distinct elements, ${\cal D}_{11} = {\cal D}_{22}$ and ${\cal D}_{33}$. Thus diffusion in the radial (perpendicular to dipole alignment) and axial (parallel to dipole alignment) directions occur with the respective different characteristic time scales
\begin{subequations}
\begin{align}
    & \tau_{r} \equiv \frac{ \sigma^2 }{ 2 {\cal D}_{11} } = \frac{128 \left(315 a^2 - 42 a a_d + 32 a_d^2\right) }{ 7875 r_\mathrm{eff}^2 } \tau_0, \\
    & \tau_{z} \equiv \frac{ \sigma^2 }{ 2 {\cal D}_{33} } = \frac{128 \left(315 a^2 + 84 a a_d + 20 a_d^2\right) }{ 7875 r_\mathrm{eff}^2 } \tau_0,
\end{align}
\end{subequations}
with $\tau_0 = \sigma^2 r_\mathrm{eff}^2 n_0 \sqrt{ \pi m \beta }$, that dictate the Gaussian hot-spot relaxation time along the radial and axial directions respectively. These time scales are of course identical in the limit of vanishing dipole moment $a_d=0$.  Their difference is quite pronounced, however, as $a_d$ increases, as illustrated in Figure \ref{fig:dispersion_timescale}. This figure uses  the experimental parameters in Tab.~\ref{tab:system_parameter}, and a hot-spot of initial width $\sigma = 5 L \approx 0.6$ (mm).  It is apparent the diffusion occurs far more rapidly in the axial direction, when dipolar scattering is significant. 

\begin{figure}[ht]
    \centering
    \includegraphics[width=\columnwidth]{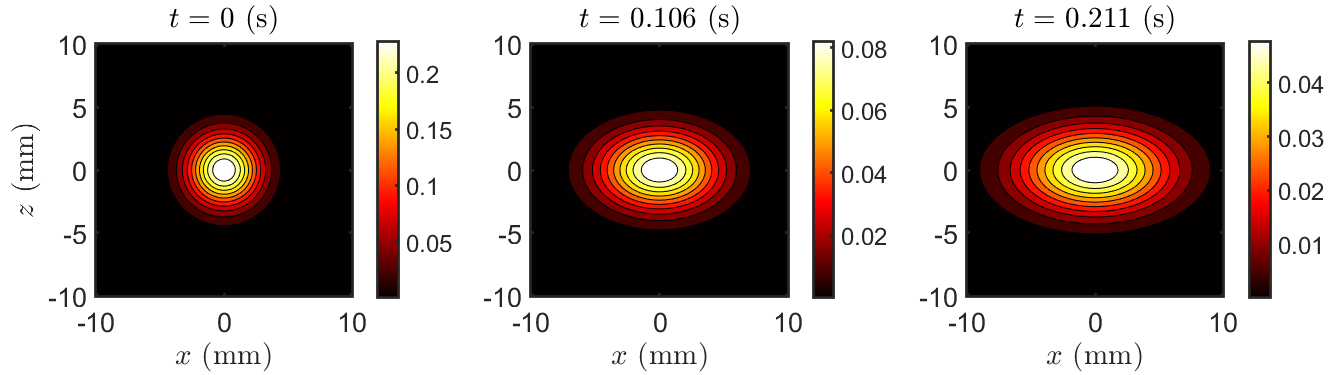}
    \caption{ Stroboscopic evolution of the temperature field variation $\epsilon(\boldsymbol{r}, t)$, at times $t = 0, 0.106, 0.211$s (plots left to right), visualized along a 2D slice in the $x,z$-plane. The initial peak temperature fluctuation amplitude is set to $\epsilon_0 = 0.25$, and the color scale for each plot is rescaled for visual clarity at each time instance. }
    \label{fig:hotspot_dispersion}
\end{figure}

With the dipoles aligned along ${\hat z}$, the explicit time evolution of the hot spot is given by
\begin{align} \label{eq:Gaussian_solution}
    \epsilon(\boldsymbol{r}, t) = \frac{ \epsilon_0 \exp\Big( -\frac{ x^2 + y^2 }{ 2 \sigma^2 \big( 1 + \frac{ t }{ \tau_r } \big) } - \frac{ z^2 }{ 2 \sigma^2 \big( 1 + \frac{ t }{ \tau_z } \big)  } \Big) }{ \sqrt{ \big( 1 + \frac{ t }{ \tau_z } \big)^2 \big( 1 + \frac{ t }{ \tau_r } \big) } }.
\end{align} Figure~\ref{fig:hotspot_dispersion} visualizes the anisotropy of thermal relaxation by showing the temperature field variation $\epsilon$ in the $x,z$-plane. We plot the time evolution of $\epsilon$ in Fig.~\ref{fig:hotspot_dispersion}, up to the geometric mean of the 2 time scales in 3 panels ($t = 0, \sqrt{\tau_r \tau_z}/2, \sqrt{\tau_r \tau_z}$), where we have set $a = 0$ to accentuate the dipolar anisotropy. With the parameters in Tab.~\ref{tab:system_parameter}, the time scales take values $\tau_r = 0.0667$s and $\tau_z = 0.667$s. The Gaussian hot-spot clearly elongates along the $x$-direction over time, demonstrating an observable effect of anisotropic thermal conductivity during thermal diffusion in the fluid.

\begin{table}[ht]
\caption{\label{tab:system_parameter} 
Table of experimental parameter values. Da $= 1.661 \times 10^{-27}$ kg stands for Dalton (atomic mass unit), $a_{0} = 5.292 \times 10^{-11}$ m is the Bohr radius and $\mu_B = 9.274 \times 10^{-24}$ J/T is the Bohr magneton. }
\begin{ruledtabular}
\begin{tabular}{l c c c}
    \multicolumn{1}{c}{\textrm{Parameter}} & \multicolumn{1}{c}{\textrm{Symbol}} & \multicolumn{1}{c}{\textrm{Value}} & \multicolumn{1}{c}{\textrm{Unit}} \\
    \colrule
    Atomic mass number, & A & 164 & Da \\
    Magnetic moment & $\mu$ & 10 & $\mu_B$ \\
    Dipole length, & $a_d$ & 199 & $a_{0}$ \\
    Equilibrium number density, & $n_0$ & $10^{13}$ & cm$^{-3}$  \\
    Equilibrium gas temperature, & $T_0$ & 300 & nK 
\end{tabular}
\end{ruledtabular}
\end{table}

\section{ \label{sec:conclusions} Discussion and Conclusion }

Normal phase gases of ultracold dipolar atoms present a vast arena for anisotropic dynamical phenomena. In large enough samples, a continuum description of these systems are warranted, permitting fluid dynamic studies. The fluid equations of motion are, however, only complete upon specification of the transport tensors, which govern the finite-time dispersive processes in the fluid. In this work, we have used the Chapman-Enskog procedure to derive analytic expressions for the anisotropic transport tensor of thermal conductivity, induced by collisions between dipolar Bosons. By construction, each tensor element is a function of the dipole-alignment angle, and functionally dependent on the ratio of dipole length to scattering length. 

We then analyzed the anisotropic effects of these thermal conductivities in the thermal relaxation of a Gaussian hot-spot, where time-dependent solutions were derived from a linearization of the viscous-free fluid equations. We find that an initially isotropic hot-spot would disperse preferentially in a direction orthogonal to the dipole orientation, opening the possibility for control of heat transport with the dipole-alignment direction. 

A comprehensive fluid description will of course require the transport tensor of viscosity to also be derived. The analytic techniques presented here permit this derivation, which will be a subject of future work. 
Another possible extension of this work is to include quantum statistical effects in computing the transport coefficients, as done in Refs.~\cite{Uehling34_APS, Nikuni98_JLTP}, but with the dipolar cross section of Ref.~\cite{Bohn14_PRA}. These effects might become relevant at temperatures closer to quantum degeneracy. Finally, we note that recent experiments have realized long-lived 3-dimensional polar molecular samples by microwave shielding \cite{Anderegg21_Sci, Schindewolf22_arxiv} or DC electric fields \cite{Li21_Nat}, promising larger and tunable electric dipole moments in collisional dipolar gases. These systems would serve as ideal platforms for experimental investigations of dipolar fluid dynamics.

\onecolumngrid

\twocolumngrid

\begin{acknowledgments}

This work is supported by the National Science Foundation under Grant Number PHY-2110327. 

\end{acknowledgments}

\appendix

\section{ \label{app:1st_order_ChapmanEnskog} The First-Order Chapman-Enskog Approximation to the Boltzmann Equation }

This appendix section details the derivation for the left-hand side of the Boltzmann equation under the Chapman-Enskog expansion to first-order \cite{Bond65_AW}. We can first write this expression as
\begin{align}
    \left( \frac{\partial }{ \partial t } + v_i \partial_i \right) f_0 &= f_0 \left( \frac{\partial }{ \partial t } + v_i \partial_i \right) \ln f_0 \nonumber\\
    &= f_0 \left( \frac{D }{ D t } + u_i \partial_i \right) \ln f_0, 
\end{align}
where we defined the material derivative 
\begin{align}
    \frac{ D }{ D t } \equiv \frac{ \partial }{ \partial t } + U_j \partial_j.  
\end{align}
From Eq.~(\ref{eq:equilibrium_ansatz}), it follows that
\begin{align}
    \ln f_0 = \frac{ 3 }{ 2 } \ln\frac{ m }{ 2\pi } + \ln ( n_0 \beta^{3/2} ) - \frac{ m \beta }{ 2 } \boldsymbol{u}^2, 
\end{align}
so 
\begin{align}
    & \left( \frac{\partial }{ \partial t } + v_i \partial_i \right) f_0 \\
    &\quad\quad = f_0 \left( \frac{D }{ D t } + u_i \partial_i \right) \left[ \ln ( n_0 \beta^{3/2} ) - \frac{ m \beta }{ 2 } \boldsymbol{u}^2 \right]. \nonumber
\end{align}
At local thermal equilibrium as prescribed by $f_0$, the equations of conservation [Eq.~(\ref{eq:continuum_conservation_laws})] reduce to
\begin{subequations}
\begin{align}
    & \left( \frac{ D }{ D t } + \partial_j U_j \right) n_0 = 0, \\
    & \left( \frac{ D }{ D t } + \partial_j U_j \right) U_i = - \frac{ 1 }{ n_0 } \partial_i \left( \frac{ n_0 }{ m \beta } \right), \\ 
    & \left( \frac{ D }{ D t } + \partial_j U_j \right) \beta = \frac{ 5 }{ 3 } \beta \partial_j U_j,
\end{align}
\end{subequations}
from which the equations of continuity and energy balance can be combined to give the relation
\begin{align}
    \frac{ D }{ D t } \ln( n_0 \beta^{3/2} ) = 0,
\end{align}
identifying the quantity $\ln( n_0 \beta^{3/2} )$ as an adiabatic invariant. This simplifies the expression to
\begin{align}
    & \left( \frac{ D }{ D t } + u_i \partial_i \right) f_0 \\
    &\quad\quad = f_0 u_j \partial_j \ln ( n_0 \beta^{3/2} ) - f_0 \left( \frac{D }{ D t } + u_i \partial_i \right) \frac{ m \beta }{ 2 } \boldsymbol{u}^2. \nonumber
\end{align}
Applying the material derivative to the term in $\boldsymbol{u}^2$ gives
\begin{align}
    \frac{ D }{ D t } \left( \frac{ m \beta }{ 2 } \boldsymbol{u}^2 \right) &= \frac{ m }{ 2 } \left( \boldsymbol{u}^2 \frac{ D \beta }{ D t } + \beta \frac{ D \boldsymbol{u}^2 }{ D t } \right) \nonumber\\
    &= m \beta \left( \frac{ 1 }{ 3 } \boldsymbol{u}^2 \partial_i U_i - u_i \frac{ D U_i }{ D t } \right) \nonumber\\
    &= m \beta \left[ \frac{ 1 }{ 3 } \boldsymbol{u}^2 \partial_i U_i + \frac{ u_i }{ n_0 } \partial_i \left( \frac{ n_0 }{ m \beta } \right) \right] \nonumber\\
    &= \frac{ 1 }{ 3 } m \beta \boldsymbol{u}^2 \partial_i U_i + u_i \partial_i \ln( n_0 T ),
\end{align}
thus the left-hand side of the Boltzmann equation  becomes 
\begin{widetext}
\begin{align}
    \left( \frac{ D }{ D t } + v_i \partial_i \right) \ln f_0
    &=
    u_i \partial_i \left( \frac{ 5 }{ 2 } \ln \beta - \frac{ m \beta }{ 2 } \boldsymbol{u}^2 \right) - \frac{ 1 }{ 3 } m \beta \boldsymbol{u}^2 \partial_i U_i \nonumber\\
    &=
    \frac{ 1 }{ \beta } \left( \frac{ 5 }{ 2 } - \frac{ m \beta }{ 2 } \boldsymbol{u}^2 \right) u_i \partial_i \beta + m \beta \left( u_i u_j \partial_j U_i - \frac{ 1 }{ 3 } \boldsymbol{u}^2 \partial_i U_i \right) \nonumber\\
    &=
    \left( \frac{ m \beta }{ 2 } \boldsymbol{u}^2 - \frac{ 5 }{ 2 } \right) u_i \partial_i ( \ln T ) + m \beta \left( u_i u_j - \frac{ 1 }{ 3 } \delta_{ij} \boldsymbol{u}^2 \right) \left( \frac{ \partial_j U_i + \partial_i U_j }{ 2 } - \frac{ 1 }{ 3 } \delta_{ij} \partial_k U_k \right)  
\end{align}
\end{widetext}
which is the form presented in Eq.~(\ref{eq:Boltzmann_LHS}) of the main text.

\section{ \label{app:collision_integral_T} Evaluation of the Collision Integral for Thermal Conduction }

The collision integral to be computed is written as
\begin{align}
    N_{i k} &= \frac{ 2 m \beta }{ 5 n_0 } \int d^3 u V_i(\boldsymbol{u}) C[f_0 V_k] \\
    &= \frac{ 2 m \beta }{ 5 n_0 } \int d^3 u \: V_i(\boldsymbol{u}) \int d^3 u_1 \abs{ \boldsymbol{u} - \boldsymbol{u}_1 } f_0(\boldsymbol{u}) f_0(\boldsymbol{u}_1) \nonumber\\
    &\quad\quad\quad\quad\quad\quad\quad\:\: \times \int d\Omega' \frac{d\sigma}{d\Omega'} \Delta V_k.
\end{align}
In considering both the thermal motion of the atoms and collisional processes, it is convenient to first define the velocities in terms of center-of-mass (COM) and relative (r) coordinates
\begin{subequations} \label{eq:COM_coordinates}
\begin{align}
    & \boldsymbol{u}_\mathrm{COM} = \frac{ \boldsymbol{u} + \boldsymbol{u}_1 }{2}, \\
    & \boldsymbol{u}_\mathrm{r} = \boldsymbol{u} - \boldsymbol{u}_1,
\end{align}
\end{subequations}
which allows the product of equilibrium distributions to be recast as
\begin{subequations}
\begin{align}
    & f_0(\boldsymbol{u})f_0(\boldsymbol{u}_1) = f_\mathrm{COM}(\boldsymbol{u}_\mathrm{COM}) f_\mathrm{r}(\boldsymbol{u}_\mathrm{r}), \\
    & f_\mathrm{COM}(\boldsymbol{u}_\mathrm{COM}) \equiv n_0 \left( \frac{ m \beta }{ \pi } \right)^{3/2} \exp\left( - m \beta u_\mathrm{COM}^2 \right), \\
    & f_\mathrm{r}(\boldsymbol{u}_\mathrm{r}) = n_0 \left( \frac{ m \beta }{ 4 \pi } \right)^{3/2} \exp\left( - \frac{ m \beta }{ 4 } u_\mathrm{r}^2 \right). \label{eq:com_p_distribution}
\end{align}
\end{subequations}
\begin{figure}[ht]
    \centering
    \includegraphics[width=0.7\columnwidth]{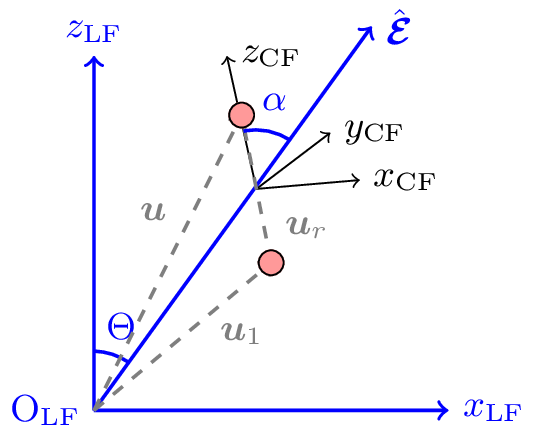}
    \caption{The collision frame (black) defined in the laboratory frame (blue) via the relative velocities between 2 colliding partners (red spheres). The angle $\alpha$ is defined as that between the vectors $\boldsymbol{u}_r$ and $\hat{\boldsymbol{{\cal E}}}$. }
    \label{fig:collision_frame_axes}
\end{figure}
Furthermore, the anisotropy of the dipolar differential cross section have us consider 2 distinct coordinate frames: 1) the laboratory-frame (LF) defined by the dipole-alignment axis $\hat{\boldsymbol{{\cal E}}}$ lying along the $x_\mathrm{LF}, z_\mathrm{LF}$-plane (Fig.~\ref{fig:lab_frame_axes}); and 2) the collision-frame (CF) defined by aligning the $\hat{z}_\mathrm{CF}$-axis to the direction of relative incoming velocities $\boldsymbol{u}_r$, for 2 colliding atoms (visualization in Fig.~\ref{fig:collision_frame_axes}). We perform the collision integral in coordinates defined with respect to the lab-frame.

To transform between coordinate frames, we construct a frame rotation matrix of direction cosines, 
\begin{align}
    R_{ \text{CF}\rightarrow\text{LF} }  &= \begin{pmatrix}
        \hat{x}_\mathrm{LF}\cdot\hat{x}_\mathrm{CF} & \hat{x}_\mathrm{LF}\cdot\hat{y}_\mathrm{CF} & \hat{x}_\mathrm{LF}\cdot\hat{z}_\mathrm{CF}  \\
        \hat{y}_\mathrm{LF}\cdot\hat{x}_\mathrm{CF} & \hat{y}_\mathrm{LF}\cdot\hat{y}_\mathrm{CF} & \hat{y}_\mathrm{LF}\cdot\hat{z}_\mathrm{CF} \\
        \hat{z}_\mathrm{LF}\cdot\hat{x}_\mathrm{CF} & \hat{z}_\mathrm{LF}\cdot\hat{y}_\mathrm{CF} & \hat{z}_\mathrm{LF}\cdot\hat{z}_\mathrm{CF}
    \end{pmatrix}, \label{eq:CF2LF_Rmatrix}
\end{align}
that takes the vector $\hat{\boldsymbol{u}}'_r$ from the CF to the LF. The differential scattering cross section is then also required to be expressed in LF coordinates during integration of the collision integral. To do so, we utilize the coordinate-independent form of the scattering amplitudes $f_B$ (for Bosons) \cite{Bohn14_PRA}
\begin{widetext}
\begin{align}
    f_B\left(\hat{\boldsymbol{u}}^{\prime}_r, \hat{\boldsymbol{u}_r}\right)
    = 
    \frac{a_d}{\sqrt{2}} \left(
        \frac{ 4 }{ 3 } - 2 a_s - \frac{ 2 ( \hat{\boldsymbol{u}_r} \cdot \hat{\mathcal{E}} )^{2} + 2 ( \hat{\boldsymbol{u}}^{\prime}_r \cdot \hat{\mathcal{E}} )^{2} - 4 ( \hat{\boldsymbol{u}_r} \cdot \hat{\mathcal{E}} ) ( \hat{\boldsymbol{u}}^{\prime}_r \cdot \hat{\mathcal{E}} )( \hat{\boldsymbol{u}_r} \cdot \hat{\boldsymbol{u}}^{\prime}_r ) }{ 1 - (\hat{\boldsymbol{u}_r} \cdot \hat{\boldsymbol{u}}^{\prime}_r )^{2} }
    \right),
\end{align}
\end{widetext}
and express that in terms of our desired coordinates which allows us to compute the differential cross section
\begin{align}
\begin{split} 
    \frac{ d \sigma }{ d \Omega' } = \abs{ f_B\left(\hat{\boldsymbol{u}}^{\prime}_r, \hat{\boldsymbol{u}_r}\right) }^2.
\end{split}
\end{align}
The above coordinate transformations are sufficient for us to now compute the collision integrals.

Expanding in terms of the COM and r coordinates of Eq.~(\ref{eq:COM_coordinates}), the collision integral becomes
\begin{align} \label{eq:Nik_collision_integral} 
    N_{i k} &= \frac{ 2 m \beta }{ 5 n_0 }
    \int d^3 {u}_\mathrm{COM} f_\mathrm{COM}(\boldsymbol{u}_\mathrm{COM}) \\
    &\quad \times \int d^3 {u}_\mathrm{r} f_\mathrm{r}(\boldsymbol{u}_\mathrm{r}) {u}_\mathrm{r} V_i(\boldsymbol{u}_\mathrm{COM}, \boldsymbol{u}_r) \int d\Omega' \frac{ d \sigma }{ d \Omega' } \Delta V_k. \nonumber
\end{align}
Collisions result in the variation
\begin{align}
    \Delta V_k &= \Delta\left[ \left( \frac{ m \beta \boldsymbol{u}^2 }{ 2 } - \frac{ 5 }{ 2 } \right) u_k \right] = \frac{ m \beta }{ 2 } \Delta ( \boldsymbol{u}^2 u_k ),
\end{align}
where the velocity terms are written in terms of CF and LF coordinates as
\begin{subequations}
\begin{align}
    & u_i = u_{\mathrm{COM},i} + \frac{1}{2} u_{\mathrm{r},i}, 
    \\
    & u_{1,i} = u_{\mathrm{COM},i} - \frac{1}{2} u_{\mathrm{r},i}, 
    \\
    & \boldsymbol{u}^2 = \boldsymbol{u}_{\mathrm{COM}}^2 + \frac{1}{4} \boldsymbol{u}_{\mathrm{r}}^2 + u_{\mathrm{COM},j} u_{\mathrm{r},j}, 
    \\
    & \boldsymbol{u}_1^2 = \boldsymbol{u}_{\mathrm{COM}}^2 + \frac{1}{4} \boldsymbol{u}_{\mathrm{r}}^2 - u_{\mathrm{COM},j} u_{\mathrm{r},j}, 
\end{align}
\end{subequations}
which gives the expansion
\begin{align}
    \Delta ( \boldsymbol{u}^2 u_i ) &= \boldsymbol{u}^{\prime 2} u'_i + \boldsymbol{u}_1^{\prime 2} u'_{1,i} - \boldsymbol{u}^2 u_i - \boldsymbol{u}_1^2 u_{1,i} \nonumber\\  
    &= \boldsymbol{u}_{\mathrm{COM}} \cdot \left( \boldsymbol{u}'_{\mathrm{r}} u'_{r,i} - \boldsymbol{u}_{\mathrm{r}} u_{r,i} \right).
\end{align} 
The integral over post-collision velocities is then performed as
\begin{align}
    \int d\Omega' \frac{ d \sigma }{ d \Omega' } \Delta V_k
    &\equiv 
    \frac{ m \beta }{ 2 }
    \int d\Omega' \frac{ d \sigma }{ d \Omega' } \Delta ( \boldsymbol{u}^2 u_k ), 
\end{align}
which when plugged back into Eq.~(\ref{eq:Nik_collision_integral}) and evaluated, gives the result of Eq.~(\ref{eq:calNij_terms}) and expressions thereafter.

\nocite{*}

\bibliography{main.bib} 

\begin{thebibliography}{37}%
\makeatletter
\providecommand \@ifxundefined [1]{%
 \@ifx{#1\undefined}
}%
\providecommand \@ifnum [1]{%
 \ifnum #1\expandafter \@firstoftwo
 \else \expandafter \@secondoftwo
 \fi
}%
\providecommand \@ifx [1]{%
 \ifx #1\expandafter \@firstoftwo
 \else \expandafter \@secondoftwo
 \fi
}%
\providecommand \natexlab [1]{#1}%
\providecommand \enquote  [1]{``#1''}%
\providecommand \bibnamefont  [1]{#1}%
\providecommand \bibfnamefont [1]{#1}%
\providecommand \citenamefont [1]{#1}%
\providecommand \href@noop [0]{\@secondoftwo}%
\providecommand \href [0]{\begingroup \@sanitize@url \@href}%
\providecommand \@href[1]{\@@startlink{#1}\@@href}%
\providecommand \@@href[1]{\endgroup#1\@@endlink}%
\providecommand \@sanitize@url [0]{\catcode `\\12\catcode `\$12\catcode
  `\&12\catcode `\#12\catcode `\^12\catcode `\_12\catcode `\%12\relax}%
\providecommand \@@startlink[1]{}%
\providecommand \@@endlink[0]{}%
\providecommand \url  [0]{\begingroup\@sanitize@url \@url }%
\providecommand \@url [1]{\endgroup\@href {#1}{\urlprefix }}%
\providecommand \urlprefix  [0]{URL }%
\providecommand \Eprint [0]{\href }%
\providecommand \doibase [0]{https://doi.org/}%
\providecommand \selectlanguage [0]{\@gobble}%
\providecommand \bibinfo  [0]{\@secondoftwo}%
\providecommand \bibfield  [0]{\@secondoftwo}%
\providecommand \translation [1]{[#1]}%
\providecommand \BibitemOpen [0]{}%
\providecommand \bibitemStop [0]{}%
\providecommand \bibitemNoStop [0]{.\EOS\space}%
\providecommand \EOS [0]{\spacefactor3000\relax}%
\providecommand \BibitemShut  [1]{\csname bibitem#1\endcsname}%
\let\auto@bib@innerbib\@empty
\bibitem [{\citenamefont {Chomaz}\ \emph {et~al.}(2022)\citenamefont {Chomaz},
  \citenamefont {Ferrier-Barbut}, \citenamefont {Ferlaino}, \citenamefont
  {Laburthe-Tolra}, \citenamefont {Lev},\ and\ \citenamefont
  {Pfau}}]{Chomaz22_arxiv}%
  \BibitemOpen
  \bibfield  {author} {\bibinfo {author} {\bibfnamefont {L.}~\bibnamefont
  {Chomaz}}, \bibinfo {author} {\bibfnamefont {I.}~\bibnamefont
  {Ferrier-Barbut}}, \bibinfo {author} {\bibfnamefont {F.}~\bibnamefont
  {Ferlaino}}, \bibinfo {author} {\bibfnamefont {B.}~\bibnamefont
  {Laburthe-Tolra}}, \bibinfo {author} {\bibfnamefont {B.~L.}\ \bibnamefont
  {Lev}},\ and\ \bibinfo {author} {\bibfnamefont {T.}~\bibnamefont {Pfau}},\
  }\href@noop {} {\bibinfo {title} {Dipolar physics: A review of experiments
  with magnetic quantum gases}} (\bibinfo {year} {2022}),\ \Eprint
  {https://arxiv.org/abs/2201.02672} {arXiv:2201.02672 [cond-mat.quant-gas]}
  \BibitemShut {NoStop}%
\bibitem [{\citenamefont {Sykes}\ and\ \citenamefont
  {Bohn}(2015)}]{Sykes15_PRA}%
  \BibitemOpen
  \bibfield  {author} {\bibinfo {author} {\bibfnamefont {A.~G.}\ \bibnamefont
  {Sykes}}\ and\ \bibinfo {author} {\bibfnamefont {J.~L.}\ \bibnamefont
  {Bohn}},\ }\href {https://doi.org/10.1103/PhysRevA.91.013625} {\bibfield
  {journal} {\bibinfo  {journal} {Phys. Rev. A}\ }\textbf {\bibinfo {volume}
  {91}},\ \bibinfo {pages} {013625} (\bibinfo {year} {2015})}\BibitemShut
  {NoStop}%
\bibitem [{\citenamefont {Tang}\ \emph {et~al.}(2016)\citenamefont {Tang},
  \citenamefont {Sykes}, \citenamefont {Burdick}, \citenamefont {DiSciacca},
  \citenamefont {Petrov},\ and\ \citenamefont {Lev}}]{Tang16_PRL}%
  \BibitemOpen
  \bibfield  {author} {\bibinfo {author} {\bibfnamefont {Y.}~\bibnamefont
  {Tang}}, \bibinfo {author} {\bibfnamefont {A.~G.}\ \bibnamefont {Sykes}},
  \bibinfo {author} {\bibfnamefont {N.~Q.}\ \bibnamefont {Burdick}}, \bibinfo
  {author} {\bibfnamefont {J.~M.}\ \bibnamefont {DiSciacca}}, \bibinfo {author}
  {\bibfnamefont {D.~S.}\ \bibnamefont {Petrov}},\ and\ \bibinfo {author}
  {\bibfnamefont {B.~L.}\ \bibnamefont {Lev}},\ }\href
  {https://doi.org/10.1103/PhysRevLett.117.155301} {\bibfield  {journal}
  {\bibinfo  {journal} {Phys. Rev. Lett.}\ }\textbf {\bibinfo {volume} {117}},\
  \bibinfo {pages} {155301} (\bibinfo {year} {2016})}\BibitemShut {NoStop}%
\bibitem [{\citenamefont {Wang}\ \emph {et~al.}(2020)\citenamefont {Wang},
  \citenamefont {Sykes},\ and\ \citenamefont {Bohn}}]{Wang20_PRA}%
  \BibitemOpen
  \bibfield  {author} {\bibinfo {author} {\bibfnamefont {R.~R.~W.}\
  \bibnamefont {Wang}}, \bibinfo {author} {\bibfnamefont {A.~G.}\ \bibnamefont
  {Sykes}},\ and\ \bibinfo {author} {\bibfnamefont {J.~L.}\ \bibnamefont
  {Bohn}},\ }\href {https://doi.org/10.1103/PhysRevA.102.033336} {\bibfield
  {journal} {\bibinfo  {journal} {Phys. Rev. A}\ }\textbf {\bibinfo {volume}
  {102}},\ \bibinfo {pages} {033336} (\bibinfo {year} {2020})}\BibitemShut
  {NoStop}%
\bibitem [{\citenamefont {Wang}\ and\ \citenamefont {Bohn}(2021)}]{Wang21_PRA}%
  \BibitemOpen
  \bibfield  {author} {\bibinfo {author} {\bibfnamefont {R.~R.~W.}\
  \bibnamefont {Wang}}\ and\ \bibinfo {author} {\bibfnamefont {J.~L.}\
  \bibnamefont {Bohn}},\ }\href {https://doi.org/10.1103/PhysRevA.103.063320}
  {\bibfield  {journal} {\bibinfo  {journal} {Phys. Rev. A}\ }\textbf {\bibinfo
  {volume} {103}},\ \bibinfo {pages} {063320} (\bibinfo {year}
  {2021})}\BibitemShut {NoStop}%
\bibitem [{\citenamefont {Tang}\ \emph {et~al.}(2015)\citenamefont {Tang},
  \citenamefont {Sykes}, \citenamefont {Burdick}, \citenamefont {Bohn},\ and\
  \citenamefont {Lev}}]{Tang15_PRA}%
  \BibitemOpen
  \bibfield  {author} {\bibinfo {author} {\bibfnamefont {Y.}~\bibnamefont
  {Tang}}, \bibinfo {author} {\bibfnamefont {A.}~\bibnamefont {Sykes}},
  \bibinfo {author} {\bibfnamefont {N.~Q.}\ \bibnamefont {Burdick}}, \bibinfo
  {author} {\bibfnamefont {J.~L.}\ \bibnamefont {Bohn}},\ and\ \bibinfo
  {author} {\bibfnamefont {B.~L.}\ \bibnamefont {Lev}},\ }\href
  {https://doi.org/10.1103/PhysRevA.92.022703} {\bibfield  {journal} {\bibinfo
  {journal} {Phys. Rev. A}\ }\textbf {\bibinfo {volume} {92}},\ \bibinfo
  {pages} {022703} (\bibinfo {year} {2015})}\BibitemShut {NoStop}%
\bibitem [{\citenamefont {Maier}\ \emph {et~al.}(2015)\citenamefont {Maier},
  \citenamefont {Ferrier-Barbut}, \citenamefont {Kadau}, \citenamefont
  {Schmitt}, \citenamefont {Wenzel}, \citenamefont {Wink}, \citenamefont
  {Pfau}, \citenamefont {Jachymski},\ and\ \citenamefont
  {Julienne}}]{Maier15_PRA}%
  \BibitemOpen
  \bibfield  {author} {\bibinfo {author} {\bibfnamefont {T.}~\bibnamefont
  {Maier}}, \bibinfo {author} {\bibfnamefont {I.}~\bibnamefont
  {Ferrier-Barbut}}, \bibinfo {author} {\bibfnamefont {H.}~\bibnamefont
  {Kadau}}, \bibinfo {author} {\bibfnamefont {M.}~\bibnamefont {Schmitt}},
  \bibinfo {author} {\bibfnamefont {M.}~\bibnamefont {Wenzel}}, \bibinfo
  {author} {\bibfnamefont {C.}~\bibnamefont {Wink}}, \bibinfo {author}
  {\bibfnamefont {T.}~\bibnamefont {Pfau}}, \bibinfo {author} {\bibfnamefont
  {K.}~\bibnamefont {Jachymski}},\ and\ \bibinfo {author} {\bibfnamefont
  {P.~S.}\ \bibnamefont {Julienne}},\ }\href
  {https://doi.org/10.1103/PhysRevA.92.060702} {\bibfield  {journal} {\bibinfo
  {journal} {Phys. Rev. A}\ }\textbf {\bibinfo {volume} {92}},\ \bibinfo
  {pages} {060702} (\bibinfo {year} {2015})}\BibitemShut {NoStop}%
\bibitem [{\citenamefont {Lucioni}\ \emph {et~al.}(2018)\citenamefont
  {Lucioni}, \citenamefont {Tanzi}, \citenamefont {Fregosi}, \citenamefont
  {Catani}, \citenamefont {Gozzini}, \citenamefont {Inguscio}, \citenamefont
  {Fioretti}, \citenamefont {Gabbanini},\ and\ \citenamefont
  {Modugno}}]{Lucioni18_PRA}%
  \BibitemOpen
  \bibfield  {author} {\bibinfo {author} {\bibfnamefont {E.}~\bibnamefont
  {Lucioni}}, \bibinfo {author} {\bibfnamefont {L.}~\bibnamefont {Tanzi}},
  \bibinfo {author} {\bibfnamefont {A.}~\bibnamefont {Fregosi}}, \bibinfo
  {author} {\bibfnamefont {J.}~\bibnamefont {Catani}}, \bibinfo {author}
  {\bibfnamefont {S.}~\bibnamefont {Gozzini}}, \bibinfo {author} {\bibfnamefont
  {M.}~\bibnamefont {Inguscio}}, \bibinfo {author} {\bibfnamefont
  {A.}~\bibnamefont {Fioretti}}, \bibinfo {author} {\bibfnamefont
  {C.}~\bibnamefont {Gabbanini}},\ and\ \bibinfo {author} {\bibfnamefont
  {G.}~\bibnamefont {Modugno}},\ }\href
  {https://doi.org/10.1103/PhysRevA.97.060701} {\bibfield  {journal} {\bibinfo
  {journal} {Phys. Rev. A}\ }\textbf {\bibinfo {volume} {97}},\ \bibinfo
  {pages} {060701} (\bibinfo {year} {2018})}\BibitemShut {NoStop}%
\bibitem [{\citenamefont {Durastante}\ \emph {et~al.}(2020)\citenamefont
  {Durastante}, \citenamefont {Politi}, \citenamefont {Sohmen}, \citenamefont
  {Ilzh\"ofer}, \citenamefont {Mark}, \citenamefont {Norcia},\ and\
  \citenamefont {Ferlaino}}]{Durastante20_PRA}%
  \BibitemOpen
  \bibfield  {author} {\bibinfo {author} {\bibfnamefont {G.}~\bibnamefont
  {Durastante}}, \bibinfo {author} {\bibfnamefont {C.}~\bibnamefont {Politi}},
  \bibinfo {author} {\bibfnamefont {M.}~\bibnamefont {Sohmen}}, \bibinfo
  {author} {\bibfnamefont {P.}~\bibnamefont {Ilzh\"ofer}}, \bibinfo {author}
  {\bibfnamefont {M.~J.}\ \bibnamefont {Mark}}, \bibinfo {author}
  {\bibfnamefont {M.~A.}\ \bibnamefont {Norcia}},\ and\ \bibinfo {author}
  {\bibfnamefont {F.}~\bibnamefont {Ferlaino}},\ }\href
  {https://doi.org/10.1103/PhysRevA.102.033330} {\bibfield  {journal} {\bibinfo
   {journal} {Phys. Rev. A}\ }\textbf {\bibinfo {volume} {102}},\ \bibinfo
  {pages} {033330} (\bibinfo {year} {2020})}\BibitemShut {NoStop}%
\bibitem [{\citenamefont {Patscheider}\ \emph {et~al.}(2021)\citenamefont
  {Patscheider}, \citenamefont {Chomaz}, \citenamefont {Natale}, \citenamefont
  {Petter}, \citenamefont {Mark}, \citenamefont {Baier}, \citenamefont {Yang},
  \citenamefont {Wang}, \citenamefont {Bohn},\ and\ \citenamefont
  {Ferlaino}}]{Patscheider21_arxiv}%
  \BibitemOpen
  \bibfield  {author} {\bibinfo {author} {\bibfnamefont {A.}~\bibnamefont
  {Patscheider}}, \bibinfo {author} {\bibfnamefont {L.}~\bibnamefont {Chomaz}},
  \bibinfo {author} {\bibfnamefont {G.}~\bibnamefont {Natale}}, \bibinfo
  {author} {\bibfnamefont {D.}~\bibnamefont {Petter}}, \bibinfo {author}
  {\bibfnamefont {M.~J.}\ \bibnamefont {Mark}}, \bibinfo {author}
  {\bibfnamefont {S.}~\bibnamefont {Baier}}, \bibinfo {author} {\bibfnamefont
  {B.}~\bibnamefont {Yang}}, \bibinfo {author} {\bibfnamefont {R.~R.~W.}\
  \bibnamefont {Wang}}, \bibinfo {author} {\bibfnamefont {J.~L.}\ \bibnamefont
  {Bohn}},\ and\ \bibinfo {author} {\bibfnamefont {F.}~\bibnamefont
  {Ferlaino}},\ }\href@noop {} {\bibinfo {title} {Accurate determination of the
  scattering length of erbium atoms}} (\bibinfo {year} {2021}),\ \Eprint
  {https://arxiv.org/abs/2112.11883} {arXiv:2112.11883 [cond-mat.quant-gas]}
  \BibitemShut {NoStop}%
\bibitem [{\citenamefont {Bohn}\ and\ \citenamefont {Jin}(2014)}]{Bohn14_PRA}%
  \BibitemOpen
  \bibfield  {author} {\bibinfo {author} {\bibfnamefont {J.~L.}\ \bibnamefont
  {Bohn}}\ and\ \bibinfo {author} {\bibfnamefont {D.~S.}\ \bibnamefont {Jin}},\
  }\href {https://doi.org/10.1103/PhysRevA.89.022702} {\bibfield  {journal}
  {\bibinfo  {journal} {Phys. Rev. A}\ }\textbf {\bibinfo {volume} {89}},\
  \bibinfo {pages} {022702} (\bibinfo {year} {2014})}\BibitemShut {NoStop}%
\bibitem [{\citenamefont {Chapman}\ and\ \citenamefont
  {Cowling}(1990)}]{Chapman90_CUP}%
  \BibitemOpen
  \bibfield  {author} {\bibinfo {author} {\bibfnamefont {S.}~\bibnamefont
  {Chapman}}\ and\ \bibinfo {author} {\bibfnamefont {T.~G.}\ \bibnamefont
  {Cowling}},\ }\href
  {https://www.cambridge.org/us/academic/subjects/mathematics/fluid-dynamics-and-solid-mechanics/mathematical-theory-non-uniform-gases-account-kinetic-theory-viscosity-thermal-conduction-and-diffusion-gases?format=PB&isbn=9780521408448}
  {\emph {\bibinfo {title} {The mathematical theory of non-uniform gases: an
  account of the kinetic theory of viscosity, thermal conduction and diffusion
  in gases}}}\ (\bibinfo  {publisher} {Cambridge university press},\ \bibinfo
  {year} {1990})\BibitemShut {NoStop}%
\bibitem [{\citenamefont {Hanley}\ \emph {et~al.}(1972)\citenamefont {Hanley},
  \citenamefont {McCarty},\ and\ \citenamefont {Cohen}}]{Hanley72_Phys}%
  \BibitemOpen
  \bibfield  {author} {\bibinfo {author} {\bibfnamefont {H.}~\bibnamefont
  {Hanley}}, \bibinfo {author} {\bibfnamefont {R.}~\bibnamefont {McCarty}},\
  and\ \bibinfo {author} {\bibfnamefont {E.}~\bibnamefont {Cohen}},\ }\href
  {https://www.sciencedirect.com/science/article/pii/0031891472901085}
  {\bibfield  {journal} {\bibinfo  {journal} {Physica}\ }\textbf {\bibinfo
  {volume} {60}},\ \bibinfo {pages} {322} (\bibinfo {year} {1972})}\BibitemShut
  {NoStop}%
\bibitem [{\citenamefont {{Mott}}\ and\ \citenamefont
  {{Massey}}(1949)}]{Mott49_OUP}%
  \BibitemOpen
  \bibfield  {author} {\bibinfo {author} {\bibfnamefont {N.~F.}\ \bibnamefont
  {{Mott}}}\ and\ \bibinfo {author} {\bibfnamefont {H.~S.~W.}\ \bibnamefont
  {{Massey}}},\ }\href
  {https://www.worldcat.org/title/theory-of-atomic-collisions/oclc/537272}
  {\emph {\bibinfo {title} {{The theory of atomic collisions. 2nd ed}}}}\
  (\bibinfo  {publisher} {Oxford University Press, Oxford},\ \bibinfo {year}
  {1949})\BibitemShut {NoStop}%
\bibitem [{\citenamefont {Dehkordi}\ \emph {et~al.}(2012)\citenamefont
  {Dehkordi}, \citenamefont {Shahzamanian}, \citenamefont {Abolhasani},\ and\
  \citenamefont {Elahi}}]{Dehkordi12_JPCS}%
  \BibitemOpen
  \bibfield  {author} {\bibinfo {author} {\bibfnamefont {M.~K.}\ \bibnamefont
  {Dehkordi}}, \bibinfo {author} {\bibfnamefont {M.}~\bibnamefont
  {Shahzamanian}}, \bibinfo {author} {\bibfnamefont {M.}~\bibnamefont
  {Abolhasani}},\ and\ \bibinfo {author} {\bibfnamefont {M.}~\bibnamefont
  {Elahi}},\ }in\ \href
  {https://iopscience.iop.org/article/10.1088/1742-6596/400/1/012029/meta}
  {\emph {\bibinfo {booktitle} {Journal of Physics: Conference Series}}},\
  Vol.\ \bibinfo {volume} {400}\ (\bibinfo {organization} {IOP Publishing},\
  \bibinfo {year} {2012})\ p.\ \bibinfo {pages} {012029}\BibitemShut {NoStop}%
\bibitem [{\citenamefont {Braginskii}(1963)}]{Braginskii63_RPP}%
  \BibitemOpen
  \bibfield  {author} {\bibinfo {author} {\bibfnamefont {S.}~\bibnamefont
  {Braginskii}},\ }\href@noop {} {\bibfield  {journal} {\bibinfo  {journal}
  {Reviews of Plasma Physics}\ }\textbf {\bibinfo {volume} {1}},\ \bibinfo
  {pages} {205} (\bibinfo {year} {1963})}\BibitemShut {NoStop}%
\bibitem [{\citenamefont {Daybelge}(1970)}]{Daybelge70_JAP}%
  \BibitemOpen
  \bibfield  {author} {\bibinfo {author} {\bibfnamefont {U.}~\bibnamefont
  {Daybelge}},\ }\href {https://doi.org/10.1063/1.1659178} {\bibfield
  {journal} {\bibinfo  {journal} {Journal of Applied Physics}\ }\textbf
  {\bibinfo {volume} {41}},\ \bibinfo {pages} {2130} (\bibinfo {year}
  {1970})}\BibitemShut {NoStop}%
\bibitem [{\citenamefont {Bruno}\ \emph {et~al.}(2006)\citenamefont {Bruno},
  \citenamefont {Catalfamo}, \citenamefont {Laricchiuta}, \citenamefont
  {Giordano},\ and\ \citenamefont {Capitelli}}]{Bruno06_POP}%
  \BibitemOpen
  \bibfield  {author} {\bibinfo {author} {\bibfnamefont {D.}~\bibnamefont
  {Bruno}}, \bibinfo {author} {\bibfnamefont {C.}~\bibnamefont {Catalfamo}},
  \bibinfo {author} {\bibfnamefont {A.}~\bibnamefont {Laricchiuta}}, \bibinfo
  {author} {\bibfnamefont {D.}~\bibnamefont {Giordano}},\ and\ \bibinfo
  {author} {\bibfnamefont {M.}~\bibnamefont {Capitelli}},\ }\href
  {https://doi.org/10.1063/1.2221675} {\bibfield  {journal} {\bibinfo
  {journal} {Physics of plasmas}\ }\textbf {\bibinfo {volume} {13}},\ \bibinfo
  {pages} {072307} (\bibinfo {year} {2006})}\BibitemShut {NoStop}%
\bibitem [{\citenamefont {De~Groot}\ and\ \citenamefont
  {Mazur}(2013)}]{deGroot13_DP}%
  \BibitemOpen
  \bibfield  {author} {\bibinfo {author} {\bibfnamefont {S.}~\bibnamefont
  {De~Groot}}\ and\ \bibinfo {author} {\bibfnamefont {P.}~\bibnamefont
  {Mazur}},\ }\href {https://store.doverpublications.com/0486647412.html}
  {\emph {\bibinfo {title} {Non-Equilibrium Thermodynamics}}},\ Dover Books on
  Physics\ (\bibinfo  {publisher} {Dover Publications},\ \bibinfo {year}
  {2013})\BibitemShut {NoStop}%
\bibitem [{\citenamefont {Saluena}\ \emph {et~al.}(1992)\citenamefont
  {Saluena}, \citenamefont {Perez-Madrid},\ and\ \citenamefont
  {Rubi}}]{Saluena92_JCP}%
  \BibitemOpen
  \bibfield  {author} {\bibinfo {author} {\bibfnamefont {C.}~\bibnamefont
  {Saluena}}, \bibinfo {author} {\bibfnamefont {A.}~\bibnamefont
  {Perez-Madrid}},\ and\ \bibinfo {author} {\bibfnamefont {J.}~\bibnamefont
  {Rubi}},\ }\href
  {https://aip.scitation.org/doi/abs/10.1063/1.462552?casa_token=O6_7RE8h8rAAAAAA:LW8mYgaJhUgCtFR32O3uQlwErGe8FdRa9T6K7gm5Q607NRN6_CmIZQlrj0IhYJKhvYpxfFHq_c4}
  {\bibfield  {journal} {\bibinfo  {journal} {The Journal of chemical physics}\
  }\textbf {\bibinfo {volume} {96}},\ \bibinfo {pages} {6950} (\bibinfo {year}
  {1992})}\BibitemShut {NoStop}%
\bibitem [{\citenamefont {Suh}\ and\ \citenamefont {Cho}(2015)}]{Suh15_MT}%
  \BibitemOpen
  \bibfield  {author} {\bibinfo {author} {\bibfnamefont {Y.~J.}\ \bibnamefont
  {Suh}}\ and\ \bibinfo {author} {\bibfnamefont {K.}~\bibnamefont {Cho}},\
  }\href@noop {} {\bibfield  {journal} {\bibinfo  {journal} {Materials
  Transactions}\ ,\ \bibinfo {pages} {M2015068}} (\bibinfo {year}
  {2015})}\BibitemShut {NoStop}%
\bibitem [{\citenamefont {Bird}\ \emph {et~al.}(2006)\citenamefont {Bird},
  \citenamefont {Stewart},\ and\ \citenamefont {Lightfoot}}]{Bird06_Wiley}%
  \BibitemOpen
  \bibfield  {author} {\bibinfo {author} {\bibfnamefont {R.~B.}\ \bibnamefont
  {Bird}}, \bibinfo {author} {\bibfnamefont {W.~E.}\ \bibnamefont {Stewart}},\
  and\ \bibinfo {author} {\bibfnamefont {E.~N.}\ \bibnamefont {Lightfoot}},\
  }\href
  {https://www.wiley.com/en-us/Transport+Phenomena\%2C+Revised+2nd+Edition-p-9780470115398}
  {\bibinfo {title} {Transport phenomena revised 2nd edition}} (\bibinfo {year}
  {2006})\BibitemShut {NoStop}%
\bibitem [{\citenamefont {Bottin}\ \emph {et~al.}(2006)\citenamefont {Bottin},
  \citenamefont {Abeele}, \citenamefont {Magin},\ and\ \citenamefont
  {Rini}}]{Bottin06_PAS}%
  \BibitemOpen
  \bibfield  {author} {\bibinfo {author} {\bibfnamefont {B.}~\bibnamefont
  {Bottin}}, \bibinfo {author} {\bibfnamefont {D.~V.}\ \bibnamefont {Abeele}},
  \bibinfo {author} {\bibfnamefont {T.~E.}\ \bibnamefont {Magin}},\ and\
  \bibinfo {author} {\bibfnamefont {P.}~\bibnamefont {Rini}},\ }\href
  {https://www.sciencedirect.com/science/article/pii/S0376042106000224}
  {\bibfield  {journal} {\bibinfo  {journal} {Progress in Aerospace Sciences}\
  }\textbf {\bibinfo {volume} {42}},\ \bibinfo {pages} {38} (\bibinfo {year}
  {2006})}\BibitemShut {NoStop}%
\bibitem [{\citenamefont {Plawsky}(2009)}]{Plawsky09_CRC}%
  \BibitemOpen
  \bibfield  {author} {\bibinfo {author} {\bibfnamefont {J.~L.}\ \bibnamefont
  {Plawsky}},\ }\href
  {https://www.routledge.com/Transport-Phenomena-Fundamentals/Plawsky/p/book/9781138080560}
  {\emph {\bibinfo {title} {Transport phenomena fundamentals}}}\ (\bibinfo
  {publisher} {CRC press},\ \bibinfo {year} {2009})\BibitemShut {NoStop}%
\bibitem [{\citenamefont {Truskey}\ \emph {et~al.}(2010)\citenamefont
  {Truskey}, \citenamefont {Yuan},\ and\ \citenamefont {Katz}}]{Truskey10_P}%
  \BibitemOpen
  \bibfield  {author} {\bibinfo {author} {\bibfnamefont {G.~A.}\ \bibnamefont
  {Truskey}}, \bibinfo {author} {\bibfnamefont {F.}~\bibnamefont {Yuan}},\ and\
  \bibinfo {author} {\bibfnamefont {D.~F.}\ \bibnamefont {Katz}},\ }\href
  {https://www.pearson.com/us/higher-education/program/Truskey-Transport-Phenomena-in-Biological-Systems-2nd-Edition/PGM78437.html}
  {\emph {\bibinfo {title} {Transport phenomena in biological systems}}}\
  (\bibinfo  {publisher} {Pearson},\ \bibinfo {year} {2010})\BibitemShut
  {NoStop}%
\bibitem [{\citenamefont {Bond}\ \emph {et~al.}(1965)\citenamefont {Bond},
  \citenamefont {Watson},\ and\ \citenamefont {Welch}}]{Bond65_AW}%
  \BibitemOpen
  \bibfield  {author} {\bibinfo {author} {\bibfnamefont {J.~W.}\ \bibnamefont
  {Bond}}, \bibinfo {author} {\bibfnamefont {K.~M.}\ \bibnamefont {Watson}},\
  and\ \bibinfo {author} {\bibfnamefont {J.~A.}\ \bibnamefont {Welch}},\
  }\href@noop {} {\emph {\bibinfo {title} {Atomic theory of gas dynamics}}},\
  Vol.\ \bibinfo {volume} {633}\ (\bibinfo  {publisher} {Addison-Wesley},\
  \bibinfo {year} {1965})\BibitemShut {NoStop}%
\bibitem [{\citenamefont {Huang}(1963)}]{Huang63_NY}%
  \BibitemOpen
  \bibfield  {author} {\bibinfo {author} {\bibfnamefont {K.}~\bibnamefont
  {Huang}},\ }\href
  {https://www.wiley.com/en-sg/Statistical+Mechanics,+2nd+Edition-p-9780471815181}
  {\emph {\bibinfo {title} {Statistical mechanics, john wily \& sons}}}\
  (\bibinfo  {publisher} {John Wiley \& Sons, Inc New York},\ \bibinfo {year}
  {1963})\ p.~\bibinfo {pages} {10}\BibitemShut {NoStop}%
\bibitem [{\citenamefont {Reif}(2009)}]{Reif09_Waveland}%
  \BibitemOpen
  \bibfield  {author} {\bibinfo {author} {\bibfnamefont {F.}~\bibnamefont
  {Reif}},\ }\href {https://www.waveland.com/browse.php?t=520} {\emph {\bibinfo
  {title} {Fundamentals of statistical and thermal physics}}}\ (\bibinfo
  {publisher} {Waveland Press},\ \bibinfo {year} {2009})\BibitemShut {NoStop}%
\bibitem [{\citenamefont {Pekeris}\ and\ \citenamefont
  {Alterman}(1957)}]{Pekeris57_PNASUSA}%
  \BibitemOpen
  \bibfield  {author} {\bibinfo {author} {\bibfnamefont {C.~L.}\ \bibnamefont
  {Pekeris}}\ and\ \bibinfo {author} {\bibfnamefont {Z.}~\bibnamefont
  {Alterman}},\ }\href {https://www.ncbi.nlm.nih.gov/pmc/articles/PMC528572/}
  {\bibfield  {journal} {\bibinfo  {journal} {Proceedings of the National
  Academy of Sciences of the United States of America}\ }\textbf {\bibinfo
  {volume} {43}},\ \bibinfo {pages} {998} (\bibinfo {year} {1957})}\BibitemShut
  {NoStop}%
\bibitem [{\citenamefont {Loyalka}\ \emph {et~al.}(2007)\citenamefont
  {Loyalka}, \citenamefont {Tipton},\ and\ \citenamefont
  {Tompson}}]{Loyalka07_PA}%
  \BibitemOpen
  \bibfield  {author} {\bibinfo {author} {\bibfnamefont {S.}~\bibnamefont
  {Loyalka}}, \bibinfo {author} {\bibfnamefont {E.}~\bibnamefont {Tipton}},\
  and\ \bibinfo {author} {\bibfnamefont {R.}~\bibnamefont {Tompson}},\ }\href
  {https://doi.org/https://doi.org/10.1016/j.physa.2006.12.001} {\bibfield
  {journal} {\bibinfo  {journal} {Physica A: Statistical Mechanics and its
  Applications}\ }\textbf {\bibinfo {volume} {379}},\ \bibinfo {pages} {417}
  (\bibinfo {year} {2007})}\BibitemShut {NoStop}%
\bibitem [{Note1()}]{Note1}%
  \BibitemOpen
  \bibinfo {note} {The coefficient $\kappa _0$ is exactly the result of Chapman
  and Enskog \cite {Chapman90_CUP}, but modified to include the quantum
  mechanical scattering length instead of a classical hardsphere
  radius.}\BibitemShut {Stop}%
\bibitem [{\citenamefont {Nikuni}\ and\ \citenamefont
  {Griffin}(1998)}]{Nikuni98_JLTP}%
  \BibitemOpen
  \bibfield  {author} {\bibinfo {author} {\bibfnamefont {T.}~\bibnamefont
  {Nikuni}}\ and\ \bibinfo {author} {\bibfnamefont {A.}~\bibnamefont
  {Griffin}},\ }\href
  {https://link.springer.com/article/10.1023/A:1022221123509} {\bibfield
  {journal} {\bibinfo  {journal} {Journal of Low Temperature Physics}\ }\textbf
  {\bibinfo {volume} {111}},\ \bibinfo {pages} {793} (\bibinfo {year}
  {1998})}\BibitemShut {NoStop}%
\bibitem [{\citenamefont {Jog}(2015)}]{Jog15_CUP}%
  \BibitemOpen
  \bibfield  {author} {\bibinfo {author} {\bibfnamefont {C.~S.}\ \bibnamefont
  {Jog}},\ }\href
  {https://www.cambridge.org/us/academic/subjects/engineering/thermal-fluids-engineering/fluid-mechanics-foundations-and-applications-mechanics-volume-2-3rd-edition?format=HB}
  {\emph {\bibinfo {title} {Fluid Mechanics: Volume 2: Foundations and
  Applications of Mechanics}}}\ (\bibinfo  {publisher} {Cambridge University
  Press},\ \bibinfo {year} {2015})\BibitemShut {NoStop}%
\bibitem [{\citenamefont {Uehling}(1934)}]{Uehling34_APS}%
  \BibitemOpen
  \bibfield  {author} {\bibinfo {author} {\bibfnamefont {E.~A.}\ \bibnamefont
  {Uehling}},\ }\href {https://doi.org/10.1103/PhysRev.46.917} {\bibfield
  {journal} {\bibinfo  {journal} {Phys. Rev.}\ }\textbf {\bibinfo {volume}
  {46}},\ \bibinfo {pages} {917} (\bibinfo {year} {1934})}\BibitemShut
  {NoStop}%
\bibitem [{\citenamefont {Anderegg}\ \emph {et~al.}(2021)\citenamefont
  {Anderegg}, \citenamefont {Burchesky}, \citenamefont {Bao}, \citenamefont
  {Yu}, \citenamefont {Karman}, \citenamefont {Chae}, \citenamefont {Ni},
  \citenamefont {Ketterle},\ and\ \citenamefont {Doyle}}]{Anderegg21_Sci}%
  \BibitemOpen
  \bibfield  {author} {\bibinfo {author} {\bibfnamefont {L.}~\bibnamefont
  {Anderegg}}, \bibinfo {author} {\bibfnamefont {S.}~\bibnamefont {Burchesky}},
  \bibinfo {author} {\bibfnamefont {Y.}~\bibnamefont {Bao}}, \bibinfo {author}
  {\bibfnamefont {S.~S.}\ \bibnamefont {Yu}}, \bibinfo {author} {\bibfnamefont
  {T.}~\bibnamefont {Karman}}, \bibinfo {author} {\bibfnamefont
  {E.}~\bibnamefont {Chae}}, \bibinfo {author} {\bibfnamefont {K.-K.}\
  \bibnamefont {Ni}}, \bibinfo {author} {\bibfnamefont {W.}~\bibnamefont
  {Ketterle}},\ and\ \bibinfo {author} {\bibfnamefont {J.~M.}\ \bibnamefont
  {Doyle}},\ }\href {https://doi.org/10.1126/science.abg9502} {\bibfield
  {journal} {\bibinfo  {journal} {Science}\ }\textbf {\bibinfo {volume}
  {373}},\ \bibinfo {pages} {779} (\bibinfo {year} {2021})}\BibitemShut
  {NoStop}%
\bibitem [{\citenamefont {Schindewolf}\ \emph {et~al.}(2022)\citenamefont
  {Schindewolf}, \citenamefont {Bause}, \citenamefont {Chen}, \citenamefont
  {Duda}, \citenamefont {Karman}, \citenamefont {Bloch},\ and\ \citenamefont
  {Luo}}]{Schindewolf22_arxiv}%
  \BibitemOpen
  \bibfield  {author} {\bibinfo {author} {\bibfnamefont {A.}~\bibnamefont
  {Schindewolf}}, \bibinfo {author} {\bibfnamefont {R.}~\bibnamefont {Bause}},
  \bibinfo {author} {\bibfnamefont {X.-Y.}\ \bibnamefont {Chen}}, \bibinfo
  {author} {\bibfnamefont {M.}~\bibnamefont {Duda}}, \bibinfo {author}
  {\bibfnamefont {T.}~\bibnamefont {Karman}}, \bibinfo {author} {\bibfnamefont
  {I.}~\bibnamefont {Bloch}},\ and\ \bibinfo {author} {\bibfnamefont {X.-Y.}\
  \bibnamefont {Luo}},\ }\href@noop {} {\bibinfo {title} {Evaporation of
  microwave-shielded polar molecules to quantum degeneracy}} (\bibinfo {year}
  {2022}),\ \Eprint {https://arxiv.org/abs/2201.05143} {arXiv:2201.05143
  [cond-mat.quant-gas]} \BibitemShut {NoStop}%
\bibitem [{\citenamefont {Li}\ \emph {et~al.}(2021)\citenamefont {Li},
  \citenamefont {Tobias}, \citenamefont {Matsuda}, \citenamefont {Miller},
  \citenamefont {Valtolina}, \citenamefont {De~Marco}, \citenamefont {Wang},
  \citenamefont {Lassabli{\`e}re}, \citenamefont {Qu{\'e}m{\'e}ner},
  \citenamefont {Bohn} \emph {et~al.}}]{Li21_Nat}%
  \BibitemOpen
  \bibfield  {author} {\bibinfo {author} {\bibfnamefont {J.-R.}\ \bibnamefont
  {Li}}, \bibinfo {author} {\bibfnamefont {W.~G.}\ \bibnamefont {Tobias}},
  \bibinfo {author} {\bibfnamefont {K.}~\bibnamefont {Matsuda}}, \bibinfo
  {author} {\bibfnamefont {C.}~\bibnamefont {Miller}}, \bibinfo {author}
  {\bibfnamefont {G.}~\bibnamefont {Valtolina}}, \bibinfo {author}
  {\bibfnamefont {L.}~\bibnamefont {De~Marco}}, \bibinfo {author}
  {\bibfnamefont {R.~R.}\ \bibnamefont {Wang}}, \bibinfo {author}
  {\bibfnamefont {L.}~\bibnamefont {Lassabli{\`e}re}}, \bibinfo {author}
  {\bibfnamefont {G.}~\bibnamefont {Qu{\'e}m{\'e}ner}}, \bibinfo {author}
  {\bibfnamefont {J.~L.}\ \bibnamefont {Bohn}}, \emph {et~al.},\ }\href
  {https://www.nature.com/articles/s41567-021-01329-6} {\bibfield  {journal}
  {\bibinfo  {journal} {Nature Physics}\ }\textbf {\bibinfo {volume} {17}},\
  \bibinfo {pages} {1144} (\bibinfo {year} {2021})}\BibitemShut {NoStop}%
\end{thebibliography}%

\end{document}